\documentclass{article}
\usepackage{spconf,amsmath,graphicx}
\usepackage{algorithm}
\usepackage{algorithmic}
\usepackage{booktabs}
\usepackage{microtype}
\usepackage{graphicx}
\usepackage{subfigure}
\usepackage{hyperref}
\usepackage{multirow}
\usepackage{color}
\usepackage{amsthm,amsmath,amssymb}
\usepackage{mathrsfs}

\title{Unidirectional Memory-Self-Attention Transducer for Online Speech Recognition}
\name{Jian Luo, Jianzong Wang*\thanks{*Corresponding author: Jianzong Wang, jzwang@188.com}, Ning Cheng, Jing Xiao}
\address{Ping An Technology (Shenzhen) Co., Ltd.}
\begin{document}
%\ninept
%
\maketitle
\begin{abstract}
Self-attention models have been successfully applied in end-to-end speech recognition systems, which greatly improve the performance of recognition accuracy. However, such attention-based models cannot be used in online speech recognition, because these models usually have to utilize a whole acoustic sequences as inputs. A common method is restricting the field of attention sights by a fixed left and right window, which makes the computation costs manageable yet also introduces performance degradation. In this paper, we propose Memory-Self-Attention (MSA), which adds history information into the Restricted-Self-Attention unit. MSA only needs localtime features as inputs, and efficiently models long temporal contexts by attending memory states. Meanwhile, recurrent neural network transducer (RNN-T) has proved to be a great approach for online ASR tasks, because the alignments of RNN-T are local and monotonic. We propose a novel network structure, called Memory-Self-Attention (MSA) Transducer. Both encoder and decoder of the MSA Transducer contain the proposed MSA unit. The experiments demonstrate that our proposed models improve WER results than Restricted-Self-Attention models by $13.5\%$ on WSJ and $7.1\%$ on SWBD datasets relatively, and without much computation costs increase.
\end{abstract}
\begin{keywords}
Speech Recognition, Self Attention, RNN Transducer
\end{keywords}
\section{Introduction}
In the past few years, models employing transformer structure have achieved state-of-art results for many tasks, such as nature language understanding, machine translation, and speech recognition. Especially in speech recognition, lots of attention-based models have proved to obtain a substantial performance improvement~\cite{Miao2019}\cite{Pham2019}\cite{Sperber2018}\cite{Chorowski2015}. For example, self-attention blocks are successfully applied in CTC-based network, SAN-CTC~\cite{salazar2019self} showed that self-attention encoder is competitive with existing end-to-end models. Speech-transformer~\cite{dong2018speech} proposed a 2D-Attention module, which computes attention weights on both time and frequency axes, in order to extract more discriminated representations of speech features. Transformer-Transducer~\cite{yeh2019transformer} used VGGNet with causal convolution as the frontend of encoder, and self-attention transducer as the network architecture. However, such attention-based models cannot be used in online speech recognition, because these models usually have to utilize a whole acoustic sequences as inputs. An additional challenge is that the computation complexity of these models increases quadratically with input sequence length, which is unacceptable for online ASR tasks. A typical solution for this challenge is restricting the field of attention sights by a fixed left and right window, which makes the computation costs manageable but also leads to performance degradation. To overcome the drawbacks of these restricted-attention model, we proposed Memory-Self-Attention (MSA) in this paper. MSA only needs localtime features as inputs, and efficiently models long temporal contexts by attending memory states. These memory states help MSA to gain better performance than window-restricted-attention models. Moreover, the computation complexity of MSA is linear with input sequence length, which is significant for online ASR tasks.

CTC~\cite{graves2006connectionist}, Transformer~\cite{vaswani2017attention}, RNN-Transducer~\cite{tian2019self}\cite{Wang2019} are most common used architectures in speech recognition~\cite{battenberg2017exploring}. CTC~\cite{sainath2015convolutional}\cite{amodei2016deep} is first widely used to end-to-end models, but CTC has a fatal drawback that every timestep is outputted independently. Therefore, it has to be optimized jointly with external language model in practice. Transformer~\cite{Zeyer2019}\cite{Li2019} is another choice by encoder-decoder infrastructure. However, the mechanism of transformer allows the model to attend anywhere in the input sequence at each timestep. Therefore, the alignments of transformer are non-local and non-monotonic. The RNN-Transducer~\cite{graves2013speech}\cite{Tsunoo2019} was proposed as an extension to CTC, which also marginalizes over all possible alignments between the input sequence and the output targets. RNN-Transducer is typically composed of an encoder, which transforms the acoustic features into high-level representations, and a decoder, which produces linguistic outputs. Previous works employed GRU or LSTM as the encoders, giving the RNN-T its name. In this paper, we explore the possibility of replacing RNN-based encoders and decoders with our proposed MSA units, which is called Memory-Self-Attention (MSA) Transducer. MSA Transducer can learn an implicit language model and therefore removes the conditional independence assumption in CTC. More importantly, the alignments of RNN-T are local and monotonic, allowing MSA Transducer for online speech recognition tasks.

In the previous works, some unidirectional neural network architectures were proposed for online ASR tasks, such as deep LCBLSTM~\cite{Xue2017Improving}, and TDNN-LSTM~\cite{Peddinti2017Low}. Daniel Povey proposed a time-restricted-attention layer for ASR~\cite{Povey2018}, and used it in LF-MMI~\cite{povey2016purely} models which are not end-to-end. Two unidirectional architectures, the time-delay LSTM (TDLSTM) and
parallel time-delayed LSTM (PTDLSTM) were presented in~\cite{Moritz2019}. Another researches are worked on local monotonic attention~\cite{Andre2019}\cite{dong2019self}. Google proposed transformer encoders with RNN-T loss~\cite{ZhangTransformer}, and they showed that limiting the left and right context of attention per-layer can obtain not bad accuracy but still have some gap between the performance of full-attention models.

%In this paper, our main contributions are as follows,
%\begin{enumerate}
%	\item First, we propose Memory-Self-Attention(MSA), which adds recurrent information into the basic restricted attention unit.
%	\item Second, MSA Transducer, a novel network structure is designed for online speech recognition, and both encoder and decoder of the network are our proposed MSA unit.
%	\item Third, we optimize the inference algorithm for MSA Transducer, which uses Trie of closed word vocabulary, to prune unnecessary paths of beam search and promote accuracy.
%\end{enumerate}

\section{Memory-Self-Attention (MSA) Transducer}
\subsection{Model Architecture}
The speech recognition tasks can be defined as a sequence to sequence problem, which takes acoustic features $X_{t}=[x_{0},x_{1},...,x_{T}]$ as the inputs, and produces predicted label sequence $Y_{u}=[y_{0},y_{1},...,y_{U}]$ as the outputs. In which, $T$ is the acoustic frames length, and $U$ is the predicted label length.
RNN-T consists of an acoustic encoder network, a separate language model named prediction network and a joint network. For online ASR, the encoder network takes acoustic features $x_{1:t}$ as the inputs and outputs encoder states $e_{t}$. Meanwhile, the decoder network takes previous predicted label $y_{1:u-1}$ recurrently, and outputs decoder states $d_{u}$. Finally, the joint network merges the encoder and decoder states together, and produces label prediction $y_{t,u}$ at each timestep $t$.

\begin{equation}
e_{t} = \mathrm{EncoderNetwork}(x_{1:t})
\end{equation}
\begin{equation}
d_{u} = \mathrm{DecoderNetwork}(y_{1:u-1})
\end{equation}
\begin{equation}
y_{t,u} = \mathrm{JointNetwork}(e_{t}, d_{u})
\end{equation}

\begin{figure}[ht]
	\vskip 0.2in
	\begin{center}
		\centerline{\includegraphics[width=\columnwidth]{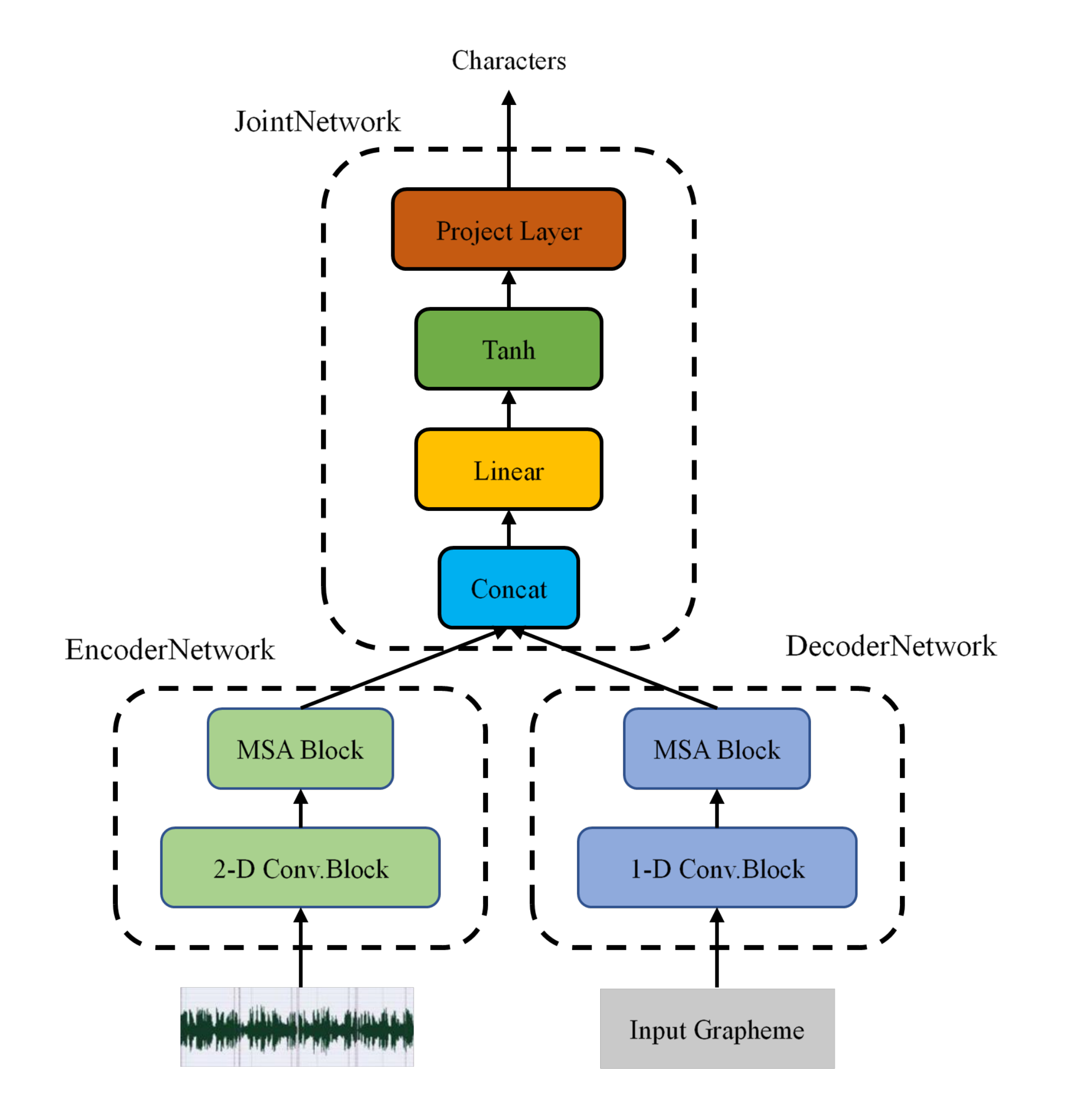}}
		\caption{Memory-Self-Attention Transducer}
		\label{fig1}
	\end{center}
	\vskip -0.2in
\end{figure}

\begin{figure}[ht]
	\vskip 0.2in
	\begin{center}
		\centerline{\includegraphics[width=\columnwidth]{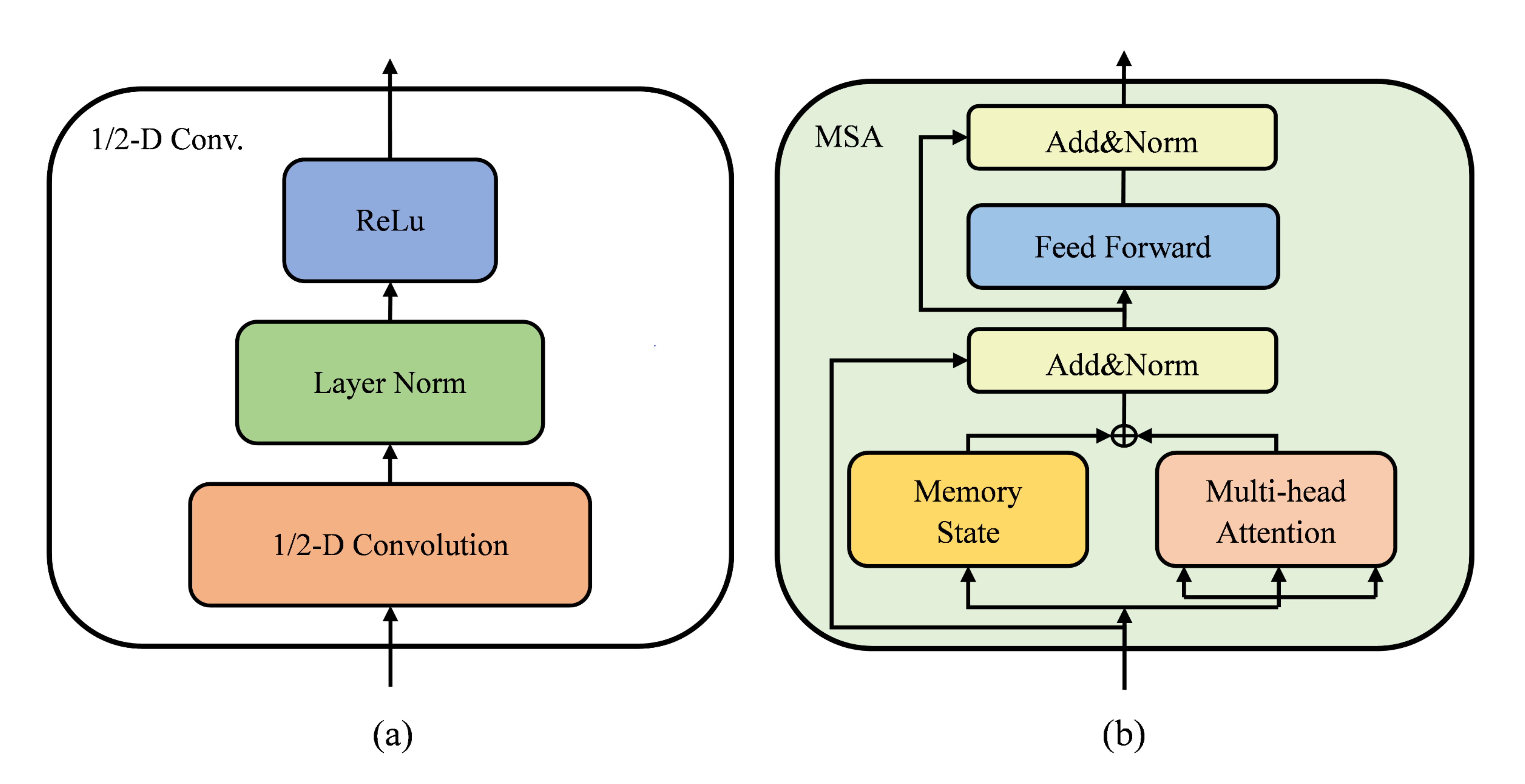}}
		\caption{(a) 1/2-D Conv Block\quad(b) MSA Block}
		\label{fig2}
	\end{center}
	\vskip -0.2in
\end{figure}

Our proposed network is based on RNN-T architecture. Encoder and decoder network contain convolutional blocks and our proposed MSA blocks, as shown in Figure~\ref{fig1}. The encoder takes acoustic features as inputs into a 2-D Conv block, in order to overcome the local variance of the features both on time and frequency axis. MSA blocks are after the 2-D Conv block, and before the joint network. The decoder also has a 1-D convolutional layer, but with grapheme inputs and convolutional kernel on time axis only. At last, a joint network is designed to concatenate both encoder and decoder output hidden states together. After linear layers and Tanh activation, the probability distribution of output labels are produced with softmax function. As shown in Figure~\ref{fig2}~a, 2-D Conv and 1-D Conv block share the similar architecture, with LayerNorm and Relu activation after convolutional layer. Moreover, the stride of 2-D convolution is set to $3$, in order to reduce the computation cost of subsequent MSA blocks.

As shown in Figure~\ref{fig2}~b, our proposed MSA block is a variant of restricted-attention unit. The first MSA block takes the output of convolutional layer as inputs, and subsequent MSA blocks are fed with the output of bottom MSA block. MSA block processes the inputs into two kinds of paths. The first path is a standard self-attention unit, and the other is a memory recurrent unit. Similar to restricted-attention unit, the MSA block takes window-restricted states into a multi-head attention layer. As $x_{t}$ is the acoustic feature of each frame, we denote the window-restricted consecutive features as $s_{t}$.
\begin{equation}
s_{t} = [x_{t-l}:x_{t+r}]
\end{equation}
\begin{equation}
m_{t} = \mathrm{MultiHeadAttention}(s_{t})
\end{equation}
\begin{equation}
h_{t} = \mathrm{LSTM}(h_{t-1}, s_{t})
\end{equation}
\begin{equation}
f_{t} = \mathrm{LayerNorm}(m_{t}+h_{t}+s_{t})
\end{equation}
\begin{equation}
e_{t} = \mathrm{LayerNorm}(\mathrm{FFN}(f_{t})+f_{t})
\end{equation}
In which, $l$ and $r$ is the left and right window size respectively. The multi-head attention layer can attend every position within the restricted window.
The output of multi-head attention layer is denoted as $m_{t}$. Another path of MSA block is a memory recurrent unit, where we denote the memory state as $h_{t}$. We use LSTM as the recurrent layer, which takes the previous hidden state $h_{t-1}$ and current input features $s_{t}$, outputing the current memory state $h_{t}$. Then, $s_{t}$, $h_{t}$, and $m_{t}$ are added together, feeding into a LayerNorm layer as $f_{t}$. At last, a feedfoward layer and another LayerNorm layer take $f_{t}$, and produce output states of encoder network as $e_{t}$.

\subsection{Complexity Analysis}
\begin{figure}[ht]
	\vskip 0.2in
	\begin{center}
		\centerline{\includegraphics[width=\columnwidth]{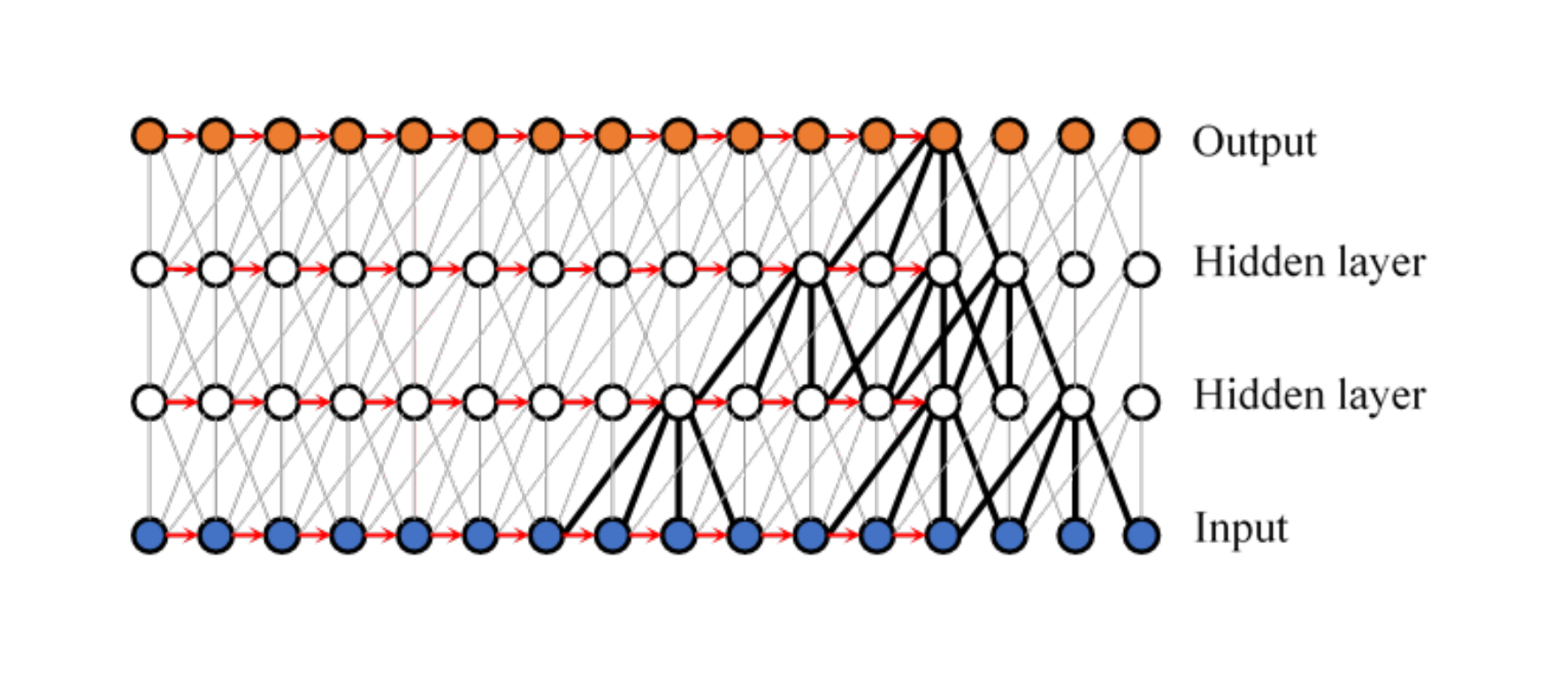}}
		\caption{Visualization of Memory-Self-Attention}
		\label{fig3}
	\end{center}
	\vskip -0.2in
\end{figure}

\begin{table*}[ht]
	\caption{\label{tab1} Operation Complexity Analysis}
	\centering	
	\begin{tabular}{p{2cm} p{4.5cm} p{3.5cm} p{2.5cm} p{3cm}}
		\toprule  %添加表格头部粗线
		
		\textbf{Model Type} & \textbf{Model Structure} & \textbf{Operations Per-layer} & \textbf{Seq-Operations} & \textbf{Max Path Length} \\
		\hline		
		Bidirectional & BLSTM  & $O(Td^2)$ & $O(T)$ & $O(T)$ \\
		Bidirectional & Unrestricted-SA & $O(T^2d)$ & $O(1)$ & $O(T)$ \\
		\hline
		Unidirectional & LSTM & $O(Td^2)$ & $O(T)$ & $O(T)$ \\
		Unidirectional & Restricted-SA (left=inf, right=r) & $O(T^2d)$ & $O(1)$ & $O(T)$ \\
		Unidirectional & Restricted-SA (left=l, right=r) & $O(T(l+r)d)$ & $O(1)$ & $O(ml+mr)$ \\
		Unidirectional & \textbf{Memory-SA} & $O(T(l+r+d)d)$ & $O(T)$ & $O(T)$ \\
		\bottomrule
	\end{tabular}	
\end{table*}

As shown in Figure~\ref{fig3}, Memory-Self-Attention not only attends the current window-restricted states but also history memory states, allowing the model to capture long-term dependency. In this section, we compare MSA with other model structure on the relationship between operation complexity and their reception of fields. As well known, BLSTM and unrestrict-self-attention are most popular used structures in acoustic network. However, such kinds of models cannot be used in online speech recognition, because these models usually have to utilize a whole acoustic sequences as inputs. It means that they cannot produce prediction labels until they see the last frame of the inputs. A common solution is making the network unidirectional like LSTM, or restricting the field of attention sights by a fixed left and right window as Restricted-Self-Attention. All of the above methods make the computation costs manageable yet also introduce performance degradation. As shown in Table~\ref{tab1}, it has proved that left-only Restrict-SA (left window = infinite) has a better performance than unidirectional LSTM. However the obvious drawback is that it makes operations complexity up to $O(T^2d)$, where $T$ is the length of input features and $d$ is the model dimension of hidden states. It means the inference speed will become slower and slower when speech frames are accumulated, which is unacceptable in online ASR tasks. Extreme left and right Restrict-SA (left=l, right=r) overcomes this drawback, and reduce the operations complexity to $O(T(l+r)d)$, but it also restricts the reception field to $O(ml+mr)$. Though it can stack multiple layers to acquire a larger reception field, where $m$ is the number of attention layers. Our proposed Memory-Self-Attention not only maintains the operations complexity to $O(T(l+r+d)d)$, but also extends the reception field to $O(T)$. In intuitive thinking, the complicated self-attention unit of MSA will model the localtime features carefully, and relative light memory unit of MSA will look as far as possible of the history information.

\section{Experiments}
\begin{table*}[t]
	\centering
	\caption{Performance of Memory-Self-Attention Model}
	\label{tab2}
	\begin{tabular}{p{2cm} p{4.5cm} p{4cm} p{1cm}p{1cm}p{1cm}p{1cm}}
		\hline
		\multirow{2}{*}{\textbf{Train Loss}} &\multirow{2}{*}{\textbf{Encoder}}&\multirow{2}{*}{\textbf{Decoder}} & \multicolumn{2}{c}{\textbf{WSJ}}&\multicolumn{2}{c}{\textbf{SWBD}}\\
		
		& & & CER & WER & CER & WER \\
		\hline
		
		CTC & LSTMx4 & --- & 6.07 & 16.76 & 17.99 & 35.43 \\
		CTC & Restricted-SAx12 (l=16, r=4) & --- & 5.21 & 14.22 & 16.42 & 33.70 \\
		CTC & \textbf{Memory-SAx12 (l=16, r=4)} & --- & \textbf{4.65} & \textbf{12.29} & \textbf{15.51} & \textbf{31.32} \\
		\hline
		
		%0.25x-----------------
		RNN-T & Memory-SAx12 (l=16, r=4) & LSTMx2 & \textbf{6.43} & \textbf{15.04} & 19.79 & 34.55 \\
		RNN-T & Memory-SAx12 (l=16, r=4) & Restricted-SAx6 (l=10, r=0) & 7.99 & 15.41 & 17.98 & 32.22 \\
		RNN-T & \textbf{Memory-SAx12 (l=16, r=4)} & \textbf{Memory-SAx2 (l=10, r=0)} & 6.98 & 15.21 & \textbf{17.32} & \textbf{31.67} \\
		\hline
	\end{tabular}
	\label{table_MAP}
\end{table*}

Our experimental works were implemented by evaluating the performance of our models on two publicly available ASR corpus (WSJ and SWBD). As a comparison, we trained the proposed MSA encoder with CTC loss, and trained the MSA Transducer with RNN-T loss. Both Character Error Rates (CER) and Word Error Rates (WER) are evaluated, and the results are summarized as Table~\ref{tab2}.

%\subsection{Dataset}
%$\textbf{Wall Street Journal(WSJ).}$ This dataset consists of read speech with texts drawn from a machine-readable corpus of Wall Street Journal news text, and contains about 80 hours speech data. All the models are trained on si284 dataset, turn the parameters on dev93, and test on eval92 dataset eventually.\\
%$\textbf{Switchboard-1 Release(SWBD)}$ This corpus contains over 300 hours of conversational telephone speech data for training. And we evaluate the model performance on HUB5 eval2000 dataset.

\subsection{Experimental Setup}
As inputs to the system, the audio data is encoded with mean-variance normalized Fbank $40$ coefficients (plus energy) of $25ms$ frame length and $10ms$ frame shift, together with their first and second temporal derivatives. Therefore, each input vector is $123$ dimensions. At each timestep, we concat the left and right $9$ frames into one feature vector, so total $19$ frames are used as the input vector. All of the speech text are capitalized, and $26$ grapheme labels (plus $-$, $'$, blank and $\emptyset$) are used during training and decoding, so the output is $30$ classes for scoring. The whole system was set up on Pytorch framework, and used $4$ Nvidia V100 cards for training. Pytorch DataParallel is used to train mini-batch data over multiple GPUs with $16$ samples per batch.

In the training stage of MSA Transducer, the model is started with $25000$ warmup steps, and we divide the learning rate by $10$ for every $20$ epochs. Adam optimizer with $\beta_1=0.9$, $\beta_2=0.98$ and $epsilon=10^{-9}$ is used. The standard MSA Transducer in most experiments has the following configuration. The encoder has (1) Two 2-D convolutional blocks, each with two convolutional layers with kernel size $=21\times5$ and channel $=32$. The second convolutional layer has stride $=3$ in temporal dimension. (2) Twelve MSA blocks, with $1024$ attention layer dimension, $8$ multi-heads attention, and $2048$ feedforward layer dimension. The decoder is similar with the encoder, but with two 1-D convolutional blocks with kernel size $=5$ and stride $=1$, followed by another two MSA blocks. Finally, the linear and projection layer dimension is $1024$, with dropout $=0.2$ for preventing overfit. In the inference stage, we used the standard beam search algorithm. More specificly, the beam width of CTC is set to $200$, and $32$ for RNN-T.

Like previous work in \cite{rao2017exploring}, we pre-trained the encoder network with CTC loss, and pre-trained the decoder network with cross-entropy loss. The joint network is fine-tuned with RNN-T loss eventually. The pre-trained encoder is connected to a $1024$ dimension feed-forward layer and a softmax layer, to output grapheme label probabilities. The performance of this pre-trained CTC model is shown in Table~\ref{tab2} as comparison. The experimental results infer that pre-training can accelerate the convergence of MSA Transducer, and it is the essential procedure to achieve better performance.

\subsection{Results}
The performance of all the models are summarized as Table~\ref{tab2}. Because our works focus on the online ASR tasks, we compare them with forward-only networks. We take $4$-layer unidirectional LSTM and $12$-layer Restricted-Self-Attention as our baseline models, where $l=16$ and $r=4$ is the left and right context window that every position in self-attention can attend. In addition, both CTC model and RNN-T model are evaluated with language model by shallow fusion for better performance. We used standard kaldi s5 recipe to train the language model. Specificly, 3-Gram model trained on the extended text data is incorporated with the CER and WER test of WSJ, and 4-Gram model trained on Switchboard and Fisher transcripts is used for SWBD test.

The experiments on encoder part of the model demonstrate that both Restricted-SA and our proposed Memory-SA can achieve better results than baseline forward-only LSTM network. The MSA model with CTC loss improves WER results than Restricted-SA network $13.5\%$ relatively on WSJ test data, and $7.1\%$ on SWBD dataset. However, the models with RNN-T loss do not beat ones with CTC loss. We thought that the decoder part of RNN-T models might not have enough data to learn a good implicit language model on small datasets. In our experiments, we also found that external language model trained on the same transcripts greatly improves the results of CTC models but has little impact on RNN-T models. 

\section{Conclusion}
In this paper, we present Memory-Self-Attention (MSA) Transducer, which adds history information into the Restricted-Self-Attention unit. We trained the MSA models on RNN-T loss, making it suitable for online ASR tasks, because the alignments of RNN-T are local and monotonic. Our experiments show that MSA has better results than basic LSTM and window-restricted-attention networks. Moreover, MSA Transducer achieves these results without much computation cost increase, because it only needs localtime features as inputs, and efficiently models long temporal contexts by attending memory states. We are interested to investigate the performance of MSA unit in various monotonic models such as truncated attention-based speech transformer. Exploring better architectures that add history information into the self-attention models, can also be extended. We will leave these ideas as future works.

\section{Acknowledgement}
This paper is supported by National Key Research and Development Program of China under grant No. 2017YFB1401202, No. 2018YFB1003500, and No. 2018YFB0204400. Corresponding author is Jianzong Wang from Ping An Technology (Shenzhen) Co., Ltd.

\clearpage
%\vfill\pagebreak

\bibliographystyle{IEEEtran}

\bibliography{ICASSP2021_MSA_Transducer}

\end{document}